\documentstyle[12pt,a4]{article}

\newcommand{\be}{\begin{equation}}
\newcommand{\ee}{\end{equation}}

\title{Central potentials and examples
of hidden algebra structure}

\author{
{\bf Y. Brihaye, N. Devaux}\\
Department of Mathematical Physics\\
University of Mons\\
Place du Parc, B-7000 Mons, Belgium\\
\and
 {\bf Piotr Kosinski
\thanks{$^{\dagger}$ Work supported by the grant
$N^o$ ERB-CIPA-93-0670 of CEC}
}\\
Department of Theoretical Physics\\
University of Lodz\\
Pomorska 149/153, 90-236 Lodz, Poland }

\begin{document}

\begin{titlepage}
\maketitle
\thispagestyle{empty}
\begin{abstract}
We propose two generalisations of the Coulomb potential
equation of quantum mechanics and investigate the occurence
of algebraic eigenfunctions for
the corresponding Scrh\"odinger equations.
Some relativistic counterparts of these problems are also
discussed.
\end{abstract}
\end{titlepage}

\section{Introduction}

Considerable attention has been always paid to the quantum-mechanical
problems which can be solved analytically.
The reason for that is twofold. First, it appeared that some,
from the physical point of view most important problems (like,
for example, the Coulomb problem), are analytically solvable.
Moreover, every such problem provides a laboratory for testing the
general ideas, approximate methods, etc.
These so-called exactly solvable models were analysed carefully and it was
revealed that there is usually (if not always) some algebraic structure
(mainly group-theoretical one) underlying the phenomenon of solvability.

It has been discovered [1-7] quite recently that there exists a class
of quantum-mechanical models, which, being generically analytically
nontractable, allow for partial determination of the spectrum once
a fine tuning of the coupling constants has been made. Such models
were named quasi exactly solvable (QES). It was also revealed that in
many cases there exists a hidden $sl_2$ structure responsible for such
a behaviour. For the first sight, the class of such models might seem
to be very restricted. However, it has been shown recently [8]
that the Coulomb correlation problem of two electrons in one external
oscillator potential belongs to this class, explaining the
existence of particular solutions to this problem [9,10].

In this note we first point out that the hidden $sl_2$ symmetry
is also responsible for the existence of exact solutions for fraction
power potentials discussed recently in Ref.[11]. We  further study
the occurence of algebraic solutions for two classes of potentials
which, generalizing those of Refs.[8,11], do not
lead to quasi-exactly solvable equations.
The discussion involves both, non-relativistic (i.e. Schr\"odinger)
as well as relativistic equations (for instance the Klein-Gordon and
the Dirac equation).

\section{Schr\"odinger equations}

The model investigated in Ref.[11] consists of a particle moving in
a central potential
\be
V(r) = {\alpha \over r^{1/2} } + {\gamma \over r}
      + {\beta \over r^{3/2}}
\ee
(actually the $\gamma$-term is absent in Ref.[11] but it can be
included without difficulty). The Schr\"odinger equation
$(\hbar  = 2m = 1)$
\be
 \lbrace {d^2 \over dr^2} + E - V(r) - {l(l+1) \over r^2}
 \rbrace \psi (r) = 0
\ee
transforms, under the following change of variable
\be
r = x^2 \quad , \quad \psi(r) = x^{2l+2} p(x) \exp(ax^2+bx)
\ee
into
\begin{eqnarray}
&\lbrace& x {d^2 \over dx^2} + (4ax^2 + 2bx + 4l + 3) {d \over dx}
\nonumber \\
&+& 4(E+a^2)x^3 + 4(ab-\alpha)x^2 \nonumber \\
&+& (b^2 + 8a(l+1)-4\gamma)x
               +(4l+3)b \rbrace p(x) = 4 \beta p(x)
\end{eqnarray}
For bound states, E is negative, so we can choose  the free
parameters $a$ and $b$ so that the terms of highest degrees
 in Eq.(4) vanish:
\be
a = - \sqrt{-E} \quad , \quad b = {- \alpha \over \sqrt{-E}}
\ee
Now, the differential operator $D$ standing on the left-hand side
of Eq.(4) preserves the space ${\cal P}_n$ of polynomials
of degree at most $n$, provided the following relation hold
\be
b^2 + 4a(2l+2+n) - 4 \gamma = 0.
\ee
It can then be expressed in terms of the generators of the
(n+1)-dimensional representation of $sl_2$
\be
 J^+ = x^2 {d \over dx} - n x \quad ,\quad
 J^0 = x {d\over dx} - {n \over 2} \quad , \quad
 J^- =  {d \over dx}
\ee
We have indeed
\be
D = J^0 J^- + 4a J^+ + 2b J^0 + (4l+3+{n\over 2}) J^- + (4l+3+n)b
\ee
Equations (5) and (6) give the eigenvalue condition
\be
(n+2l+2)(\sqrt{-E})^3 + \gamma (\sqrt{-E})^2 = {\alpha^2 \over 4}
\ee
which has  always one positive solution for $\sqrt{-E}$.
In the particular case $\gamma = 0$ we obtain Eq.(12) of Ref.[11].

Let us fix once forever $\alpha$ and $\beta$ and choose some
natural number $n$. Then Eq.(9) fixes uniquely the energy as
a function of $\alpha,\gamma$ and $n$. At the same time,
Eq.(4) becomes an algebraic equation for $\beta$ in the
space ${\cal P}_n$. If we invoke the theorem relating the level
number to the number of modes of the corresponding eigenfunction,
then the following picture emerges. Given $\alpha, \gamma$ and $n$
and the corresponding energy $E(\alpha, \gamma, n)$ we have a set
of Hamiltonians parametrized by $\beta_i (i= 0,1,2,\cdots,n$)
such that this energy corresponds to the $i$-th eigenstate
of the Hamiltonian at $i$-th value of $\beta$. All relations
given between the coupling constants $\alpha,\beta$ in Ref.[11]
can be recovered in this way. This is exactly the second
form of quasi exactly solvable Hamiltonians in the terminology
of Turbiner [3,8].

\subsection{\bf General fractional power potentials}

Let us now consider the following generalisation of Eqs.(1),(2).
Take a positive natural N and put
\be
V(r) = \sum_{k=1}^{2N-1} {\alpha_k \over r^{k/N}}
\ee
The Schr\"odinger equation (2) is now rewritten
in terms of $x = r^{1/N}$ and of $p(x)$ given by
\be \psi(r) = x^{N(l+1)} (\exp q(x)) p(x) \quad , \quad
    q(x) = \sum_{k=1}^N \beta_k x^k
\ee
and reads
\begin{eqnarray}
&\lbrace&
  {x \over N^2} {d^2 \over dx^2}
  + [{1-N \over N^2} + {2x q' \over N^2} + {2(l+1)\over N}] {d \over dx}
\nonumber \\
& + &[E x^{2N-1} - \sum_{k=1}^{2N-1} \alpha_k x^{2N-1-k} \nonumber \\
& + & {(1-N) q' \over N^2}  + {(q'^2 + q'')x \over N^2}
    + {2(l+1) q' \over N}  ] \ \rbrace \ p(x) = 0
\end{eqnarray}
It is immediate to see that, for $N>2$, the operator, say $D$,
 on the left-hand
side is not expressible in terms of the generators (7); for instance,
the terms of the form  $x^s d_x - c x^{s-1}$ do not preserve
${\cal P}_n$ if $s>2$.
 However this does not constitute a limitation
since  (it was the case also with Eq.(4))
the eigenenergy $E$ does not appear
as the spectral parameter of Eq.(12).

The construction of polynomial solutions to Eq.(12) is
performed in a few steps. We first annihilate the pure
power terms in the operator $D$, (i.e. the terms proportional
 to $x^{2N-1},x^{2N-2},\cdots x^{N}$) by suitably choosing the
parameters $\beta_k$ appearing in the exponential prefactor.
This step is done independently of the degree of p(x).

Eq.(12) can then be rewritten in the form
\be
D p(x) = \lbrace (c_{-1} x d_x +  c'_{-1}) d_x
                  + \sum_{k=0}^{N-1} x^k (c_k x d_x + c'_k) \rbrace p(x)
=0
\ee
where $c_k,c'_k$ are constants. Let p(x) be a polynomial of degree $n$;
the equation above leads to a system of $N+n$ homogeneous linear
equations in the $n+1$ coefficients of p(x). Practically, one first
solve the sub-system of $n+1$ equations corresponding to the $n+1$
lowest powers of x. This determines, in principle, $n+1$ values for
$\alpha_{2N-1}$ (this coupling constant plays a role of the eigenvalue
of the system) and $n+1$ polynomial solutions for p(x).
The $N-1$ remaining equations are finally fulfilled by imposing
constraints among the coupling constants and the energy $E$.

In summary there are $N$ relations (including the one fixing
$\alpha_{2N-1}$) among the $2N$ parameters $E, \alpha_k$.
One of the constraints fixes $E$ in terms of the coupling constants and
the $N-1$ remaining ones define an $N$-dimensional manifold in the
$2N-1$ dimensional space of the coupling constants. This manifold
can be parametrized in terms of $\alpha_1, \cdots, \alpha_N$ or,
alternatively, in terms of the parameters entering in q(x).

The form of the constrained potentials reads trivially from Eq.(12)
if we choose p(x) as a constant. In this case, we obtain a family of
models having the same ground state energy and wave function.

\subsection{\bf General polynomial potentials}

The reasoning developed above can be repeated, mutatis mutandis,
for the generalisation of quasi exactly solvable models considered
in Ref.[8]. The generalised potential is of the form
\be
V(r) = {\alpha \over r} + \sum_{k=1}^{2N} \alpha_k r^k
\ee
and the counterpart of Eq.(12) reads
\newpage
\begin{eqnarray}
&\lbrace &x{d^2 \over dx^2} + 2(xq'+(l+1)) {d\over dx} \nonumber \\
&+& Ex - x V(x) + x(q''+q'^2) + 2 (l+1)q' \ \rbrace \ p(r) = 0
\end{eqnarray}
This equation is not quasi exactly solvable for $N>1$. However a set of
algebraic solutions can be found following the lines discussed above.

In this case there are alternative circumstances under which
the Schr\"odinger equation under consideration admits
different types of algebraic solutions.
First remark that the form (15)
of the equation is obtained after multiplication
by a  power of $x$ which makes that the coefficients
standing in front of the derivatives are polynomials.
Now, let  the "odd`` coefficients (i.e. the  $\alpha, \alpha_{2j+1}$'s)
be zero in the potential (14) and divide
the full equation (15) by $x$. If the integer N
in V(x) is of the form 2j-1, then it appears that the
polynomial q(x) is even also and so will be the full operator $D/x$.
Accordingly, its eigenfunctions are even or odd functions of $x$.
If we focus on the even solutions,
one observes that all the (apparent) singular terms involving ($1/x$)
naturally  cancel and we conclude that new
polynomial solutions exist in these cases too.

We notice that the  Coulomb interaction is now
absent and that the energy parameter $E$  plays
its role of eigenvalue of the operator $D/x$.
The degree of the potential is of the form $4j-2$. The case
$j=1$ corresponds to the harmonic oscillator (quadratic potential)
which can be solved exactly. The case $j=2$ corresponds to the famous
example of quasi exactly solvable
system (Refs.[3]) with a potential of degree six.
 It admits a total of $N+1$ algebraic solutions if the
coupling constants are suitably tuned.
The potentials of higher degree, corresponding to $j>2$, have a less
rich set of algebraic solutions, only a single  eigenvector is
available if the coupling constants fulfil all the constraints.

Finally let us note that the quasi exactly solvable central potentials
remain quasi exactly solvable if one adds a Dirac monopole
placed  at the origin. This conclusion follows immediately from
the fact that the monopole interaction contributes only to the
centrifugal part of the radial Schr\"odinger equation [12]

\section{Relativistic equations}
In the previous section, we focused our attention on the
algebraic solutions of Schr\"odinger equations. The techniques
employed can be tentatively applied in the study of the
relativistic counterparts of the Schr\"odinger equation,
for instance the Klein-Gordon and the Dirac equations treated
in the background of some radial potential $A_0 = V(r)$.

The separation of the angular variable is effective for these
equations too; in the case of the Klein-Gordon equation
the condition which determines the radial part of the wave
function reads
\be
[ {d^2 \over dr^2} - {l(l+1) \over r^2} - m^2 + (E+V(r))^2] R(r) = 0
\ee
If the potential is of the form  Coulomb + polynomial
\be
    V(r) = {\alpha \over r} + \alpha_1 r + \alpha_2 r^2
                            + \cdots + \alpha_n r^n
\ee
Eq.(16) becomes
\be
[ {d^2 \over dr^2} - { l(l+1) - \alpha^2 \over r^2}
  + {2 \alpha E \over r} + (E^2 - m^2 + 2 \alpha \alpha_1)
  + 2 ( \alpha \alpha_2 + E \alpha_1) r + \cdots + \alpha_n^2 r^{2n}]
R(r) = 0
\ee
for which the above procedure can be applied.
The value of $\mu$ in the exponential prefactor is slighly affected
by a term proportional to $\alpha^2$ and the polynomial  $q(r)$
of the exponential prefactor is of degree  $n+1$.

\noindent{\bf The Dirac equation}

After the separation of the angular variable, the radial
part of the Dirac equation reads [13]
\be
\left(  \begin{array}{c c}
         {d \over dr} - {\kappa \over r} & m - V(r)- E \\
         m+V(r)+E     & {d \over dr} + {\kappa \over r}
         \end{array} \right)
\left(   \begin{array}{c}
          f(r) \\ g(r)
         \end{array} \right) =
\left(   \begin{array}{c}
          0 \\ 0
         \end{array} \right)
\ee
where $\kappa$ is the eigenvalue of the operator $1 + L\cdot \sigma$.

For the Coulomb potential ($V(r) = \alpha /r$),
it is well known that the bound states of Eq.(19)
can be  obtained by solving algebraic equations.
We will indicate how the operator defining Eq.(19) can
be set in a form that preserves the vector
space P(n-1,n), i.e. the space
of couples of polynomials of respective degrees n-1 and n.

We first extract the standard prefactors from f(r) and g(r):
\be   f(r) = p(r) r^{\mu} \exp - \lambda r \quad , \quad
      g(r) = q(r) r^{\mu} \exp - \lambda r
\ee
and fix them by imposing $\lambda = \sqrt{m^2- E^2}$,
$\mu = \sqrt{\kappa^2 - \lambda^2}$, so that polynomial solutions
for p(r), q(r) are possible.
The differential operator acting on $p,q$ reads now
\be
D \left(   \begin{array}{c}
          p(r) \\ q(r)
         \end{array} \right) \equiv
\left(  \begin{array}{c c}
         r{d \over dr} - \lambda r +  (\mu - \kappa) & r(m-V(r)- E) \\
         r(m+V(r)+E)     &r{d \over dr} - \lambda r + (\mu + \kappa)
         \end{array} \right)
\left(   \begin{array}{c}
          p(r) \\ q(r)
         \end{array} \right) =
\left(   \begin{array}{c}
          0 \\ 0
         \end{array} \right)
\ee
where, along with Ref.[8], we multiplied the equation by r.
Let us now multiply the operator $D$  by $A U^{-1}$ on the left and
by $U B$ on the right, with
\be
A = \left(  \begin{array}{c c}
         1 & 0 \\
         -1 &1
         \end{array} \right) \quad , \quad
B = \left(  \begin{array}{c c}
         1 & -1 \\
         1 &  0
         \end{array} \right) \quad , \quad
U = \left(  \begin{array}{c c}
         \sqrt{m-E}  & \sqrt{m-E} \\
         \sqrt{m+E}  & -\sqrt{m+E}
         \end{array} \right)
\ee
the matrix $U$ is nothing but the matrix which diagonalises
the piece of the operator $D$ linear in $r$.
The new operator, say $D'$ then reads
\begin{eqnarray}
D' & = &   -2 \sqrt{m^2-E^2}
       \left(  \begin{array}{c c}
         0 & 0 \\
         r & 0
         \end{array} \right)
  - 2 \alpha  \sqrt{ {m-E \over m+E}}
       \left(  \begin{array}{c c}
         0 & 0 \\
         1 & 0
         \end{array} \right)
  -  \left(  \begin{array}{c c}
         0 & r{d\over dr}-n \\
         0 & 0
         \end{array} \right) \\
&+& \left( \begin{array}{c c}
    r {d \over dr} + \mu - \kappa + \alpha \sqrt{{ m-E \over m+E}} & 0 \\
    0 & r {d \over dr} + \mu + \kappa + \alpha \sqrt{{ m-E \over m+E}}
           \end{array} \right)
\end{eqnarray}
where we have posed
\be
  n \equiv  -\mu + { \alpha E \over \sqrt{{ m^2 - E^2}}}
\ee

The condition that n is an integer is equivalent
to the well known quantization formula of the bound energies
of the relativistic hydrogen atom.
In this cases we see that the operator $D'$ manifestly
preserves the space P(n-1,n).
It is known [5,6] that the operators preserving
this vector space constitute a projectivised  representation
of the envelopping algebra of the supersymmetric algebra osp(2,2).
The  form (23) therefore reveals a hidden symetry of
the radial Dirac equation; it can be formulated
in terms of the generators of the  super-algebra  osp(2,2).

Let us finally mention that
we could not construct any algebraic solutions of Eq.(19)
by modifying the Coulomb potential along the same
lines as in the previous section.

\vfill \eject

\end{document}